\documentclass[12pt]{article}
\usepackage{geometry}             
\usepackage{graphicx}
\usepackage{amssymb}
\usepackage{amsmath}
\usepackage{epstopdf}
\usepackage{comment}
\usepackage{cite}
\usepackage{caption}

\usepackage[hyperindex=true,
          pdfstartview=FitH,
          bookmarksnumbered=true,
          bookmarksopen=true,
          citecolor=blue,
          linkcolor=blue,
          colorlinks=true,
          pdfborder=001,
          unicode]{hyperref}

\allowdisplaybreaks

\begin{document}

\title{Charged black hole with a scalar hair \\
in $(2+1)$ dimensions}
\author{Wei Xu and 
Liu Zhao\thanks{Correspondance author}\\
School of Physics, Nankai University, 
Tianjin 300071, China\\
{\em email}: 
\href{mailto:xuweifuture@mail.nankai.edu.cn}{xuweifuture@mail.nankai.edu.cn} 
and 
\href{mailto:lzhao@nankai.edu.cn}{lzhao@nankai.edu.cn}}
\date{}
\maketitle

\begin{abstract}
We obtain and analyze an exact solution to Einstein-Maxwell-scalar theory in 
$(2+1)$ dimensions, in which the scalar field couples to gravity in a 
nonminimal 
way, and it also couples to itself with the self-interacting potential
solely determined by the metric ansatz. A negative cosmological constant 
naturally emerges as a constant term in the scalar potential. The metric is static and 
circularly symmetric and contains a curvature singularity at the origin. The 
conditions for the metric to contain 0, 1 and 2 horizons are identified, and the effects
of the scalar and electric charges on the size of the black hole radius are 
discussed. Under proper choices 
of parameters, the metric degenerates into some previously known solutions
in $(2+1)$-dimensional gravity.
\vspace{3mm}

\noindent Keywords: charged hairy black hole, $(2+1)$-dimensional gravity, 
non-minimal coupling, self-interacting potential
\vspace{3mm}

\noindent PACS: 04.20.Jb, 04.40.Nr, 04.70.-s

\end{abstract}

\section{Introduction}

Gravity in $(2+1)$-dimensional spacetime has been a fascinating area of 
theoretical investigations during the last few decades. Such studies were 
initiated in the early 1980s \cite{JK,JK2,Jackiw,Giddings}. 
It was once believed 
\cite{Giddings} that there is no black hole solutions in
$(2+1)$ dimensions in the absence of a matter source, because there is no propagating
degrees of freedom. However, since the discovery of 
Ba\~nados-Teitelboim-Zanelli (BTZ) \cite{Banados:1992wn} and 
Martinez-Teitelboim-Zanelli (MTZ) \cite{Martinez:1999p2523}  black holes
and the asymptotic conformal symmetry
\cite{Brown:1986p547,Henneaux:2002p2538}, 
it became increasingly clear that gravity in
$(2+1)$ dimensions is much more interesting in its own right, not only because
black hole solutions exist but also because such theories are ideal theoretical
laboratories for studying AdS/CFT, AdS/CMT (i.e., condensed matter theory) \cite{Brown:1986p547,Henneaux:2002p2538,
Hasanpour:2011p2274,Martinez:1996p2505,Natsuume:1999p2556,
Correa:2010p2540,Correa:2011p2543,Correa:2012p2547,
Blagojevic:2013p2533,Chen:2013p2529,
Zeng:2009p2536,Colgain:2010p108},
gravity-fluid dual \cite{Cai:2012vr}, (holographic) phase transitions
\cite{Myung:2008p2548,Eune:2013p2524},
etc.
Moreover, the study of gravity in $(2+1)$ dimensions is also expected to shed some light
on the understanding of more realistic or complicated cases of four- and higher
dimensional gravities.

Recently, $(2+1)$-dimensional gravity with a matter source has attracted 
considerable interest. 
Besides the standard Maxwell source\cite{Martinez:1999p2523,
Kamata:1995p2516,Martinez:2006p2553}, 
the inclusion of extra scalar field(s)
\cite{Henneaux:2002p2538, 
Chan:1996p2544,Abramo:2003p2564,Nadalini:2007p2561,Acena:2012p2551,
Tafel:2011p2531,
Pugliese:2013p2525,Aparicio:2012p2461,Dias:2011p376,Bekenstein:1996p2554,
Correa:2010p2540,Correa:2011p2543,Correa:2012p2547,
Cadoni:2011p911,Kolyvaris:2009p2549,Degura:1998p2546,
Banados:2005p2542,BerredoPeixoto:2002p2539,Schmidt:2012p2518,
Radu:2005p2569,Martinez:1996p2505,Zeng:2009p2536,
Kwon:2012p2496,Martinez:2002p2565,
Jamil:2012p2563,Kuriakose:2008p2559,
Dotti:2007p2555,Kolyvaris:2011p1518,Myung:2008p2548,Martinez:2006p2553,
Martinez:2004p2545
}, higher rank tensor fields 
\cite{Anninos:2011p953,Perez:2013p2534,Chen:2012p2513,
David:2012p2251,
Chen:2012p2249,Ammon:2011p942}, higher curvature terms
\cite{Bergshoeff:2009p312,Gabadadze:2012p2512,
Hohm:2012p2150,Chen:2012p1989,Ohta:2011p1120,Bagchi:2011p850,
Song:2008p316,Oliva:2012p2290}, and/or gravitational 
Chern-Simons terms \cite{JK,JK2,Brihaye:2010p17} are also 
intensively studied. Unlike gravities in four- and higher 
dimensions, it is possible to include a finite number of higher rank tensor 
fields in $(2+1)$ dimensions \cite{Anninos:2011p953,Perez:2013p2534,
Chen:2012p2513,David:2012p2251,
Chen:2012p2249,Ammon:2011p942}. The inclusion of 
gravitational Chern-Simons terms 
will bring in propagating degrees of freedom in $(2+1)$ dimensions 
\cite{Moore:1989yh,Elitzur:1989nr}. 
Moreover, 
it is often much easier to obtain and analyze black hole solutions in $(2+1)$
dimensions than in other dimensions.  

In this paper, we aim to study the black hole solution in an 
Einstein-Maxwell-scalar gravity with a nonminimally coupled scalar field in $(2+1)$ 
dimensions. Gravity coupled with a scalar field is not a new idea. Black hole solutions
in such theories are known as hairy black holes,  and there is already a huge 
amount of literature on this subject  
\cite{Henneaux:2002p2538, 
Chan:1996p2544,Abramo:2003p2564,Nadalini:2007p2561,Acena:2012p2551,
Brihaye:2011p2541,Tafel:2011p2531,
Pugliese:2013p2525,Aparicio:2012p2461,Dias:2011p376,Bekenstein:1996p2554,
Correa:2010p2540,Correa:2011p2543,Correa:2012p2547,
Cadoni:2011p911,Kolyvaris:2009p2549,Degura:1998p2546,
Banados:2005p2542,BerredoPeixoto:2002p2539,Schmidt:2012p2518,
Anabalon:2009p2560,Anabalon:2012p2557,Anabalon:2012p2562,
Martinez:1996p2505,Radu:2005p2569,Martinez:2005p2568,Zeng:2009p2536,
Nucamendi:2003p2566,Kwon:2012p2496,Martinez:2002p2565,
Jamil:2012p2563,Kuriakose:2008p2559,Charmousis:2009p2558,
Dotti:2007p2555,Kolyvaris:2011p1518,Myung:2008p2548,Martinez:2006p2553,
Martinez:2004p2545
}, and the spacetime  
is not only restricted to be $(2+1)$-dimensional \cite{Brihaye:2011p2541,
Anabalon:2009p2560,Anabalon:2012p2557,Anabalon:2012p2562,Martinez:2005p2568,
Nucamendi:2003p2566,Charmousis:2009p2558}. 
The scalar field $\phi$ may be coupled 
either minimally \cite{Correa:2010p2540,Correa:2011p2543,Correa:2012p2547,
Henneaux:2002p2538,Martinez:2006p2553,Martinez:2004p2545,
Cadoni:2011p911,Kolyvaris:2009p2549,Degura:1998p2546,Banados:2005p2542,
BerredoPeixoto:2002p2539,Schmidt:2012p2518,Kwon:2012p2496}
or nonminimally 
\cite{Hasanpour:2011p2274,
Natsuume:1999p2556,Martinez:1996p2505,
Anabalon:2009p2560,Anabalon:2012p2557,
Anabalon:2012p2562,
Radu:2005p2569,Martinez:2005p2568,Nucamendi:2003p2566,Martinez:2002p2565,
Jamil:2012p2563,Kuriakose:2008p2559,Charmousis:2009p2558,
Dotti:2007p2555,Kolyvaris:2011p1518} to gravity, and it may 
or may not couple to itself through a self-interacting potential $U(\phi)$. 
In the model with which
we shall be dealing, $\phi$ couples to gravity in a nonminimal 
way, and it also couples to itself via a self-potential $V(\phi)$. The action reads
\begin{align}
I=\frac{1}{2}\int\mathrm{d}^3 x\sqrt{-g}\left[R
-g^{\mu\nu}\nabla_{\mu}\phi\nabla_{\nu}\phi
-\xi R\phi^2 
-2V(\phi)
-\frac{ 1}{4}F_{\mu\nu}F^{\mu\nu}\right],
\label{action}
\end{align}
where $\xi$ is a constant signifying the coupling strength between gravity and the scalar 
field, and we have set the gravitational constant $\kappa$ equal to unity. 
A similar action in four-dimensional spacetime was studied in 
Ref.\cite{Martinez:2005p2568}.

In the absence of the scalar potential $V(\phi)$ and the Maxwell field, the constant
value $\xi=\frac{1}{8}$ will make the coupling between gravity and the free scalar
field $\phi$ conformally invariant. In the presence of $V(\phi)$, however, the conformal
symmetry will, in general, be broken. Nonetheless, the special value $\xi=\frac{1}{8}$
for the coupling constant will greatly simplify the solution. So, we will stick to this
particular value of the gravity-scalar coupling. Notice that we did not include 
explicitly a cosmological constant term in the action; however, it will turn 
out that, as far as a black hole solution is concerned, a negative cosmological constant
will automatically emerge.  

The rest of the paper is organized as follows. In Sec. 2 we describe the exact 
solution
to the field equations, which follow from the action (\ref{action}) as well as the 
associated scalar potential. Some of the basic properties of the scalar potential
are discussed. Meanwhile the basic geometric properties of the metric are also 
outlined. In Sec. 3 we describe some special degenerated cases of the solution, 
some of which are already known in the literature. Sec. 4 is devoted to the 
analysis of the metric, in particular, the conditions for the metric to behave as an 
asymptotic AdS${}_{3}$ spacetime with a naked singularity, as an extremal charged 
hairy black hole and as a non-extremal charged hairy black hole are identified. 
In Sec. 5, we discuss the effect of the scalar and electric charges on the size 
of the black hole horizons. Finally, in Sec. 6
we make some discussions and outline some of the open problems that we intend to 
solve in subsequent investigations.

\section{Scalar potential and the static, circularly symmetric solution}  

By a straightforward variational process and discarding all possible boundary terms, we 
can write down the field equations 
associated with the action (\ref{action}) as follows,
\begin{align}
  &G_{\mu\nu}-T^{[\phi]}_{\mu\nu}-T^{[A]}_{\mu\nu}
  +V(\phi)g_{\mu\nu}=0,\label{Einstein}\\
  &\square\phi-\xi R\phi-V_{\phi}=0,\label{scalar}\\
  &\partial_{\nu}\big(\sqrt{-g}F^{\mu\nu}\big)=0,\label{Maxwell}
\end{align}
where
\begin{align*}
  V_{\phi}&=\partial_{\phi}V(\phi),\\
  \square&\equiv g^{\mu\nu}\nabla_{\mu}\nabla_{\nu},\\
  T^{[\phi]}_{\mu\nu}&=\partial_{\mu}\phi\partial_{\nu}\phi
  -\frac{1}{2}g_{\mu\nu}\nabla^{\rho}\phi\nabla_{\rho}\phi
  +\xi\big(g_{\mu\nu}\square-\nabla_{\mu}\nabla_{\nu}+G_{\mu\nu}
  \big)\phi^2,\\
  T^{[A]}_{\mu\nu}&=\frac{ 1}{2}\bigg(F_{\mu \rho}F_{\nu}{}^{\rho}
  -\frac{1}{4}g_{\mu\nu}F_{\rho\sigma}F^{\rho\sigma}\bigg).
\end{align*}
As mentioned earlier, we take $\xi=\frac{1}{8}$ throughout this paper. Note that we 
did not explicitly specify the scalar potential. Actually, it will be determined uniquely
by the form of the metric ansatz to be given below. The same phenomenon also happens 
in the study of four-dimensional hairy black holes 
\cite{Anabalon:2009p2560,Anabalon:2012p2557,Anabalon:2012p2562}. 

\subsection{Circularly symmetric solution}

We are interested in static, circularly symmetric solutions. To obtain such a solution, 
we assume that the metric takes the following form,
\begin{align}
  \mathrm{d} s^2=-f(r)\mathrm{d} t^2+\frac{1}{f(r)}\mathrm{d} r^2+r^2\mathrm{d} 
  \psi^2, \label{mt}
\end{align}
where the coordinate ranges are given by $-\infty<t<\infty$, $r\geq0$, and 
$-\pi\leq\psi\leq\pi$. We also assume that both the scalar field $\phi$ and the 
Maxwell field $A_\mu$ depend only on the radial coordinate $r$. Under such 
assumptions, the Maxwell equation (\ref{Maxwell}) gives
\begin{align}
   A_{\mu}dx^\mu=-Q\ln\bigg(\frac{r}{r_0}\bigg)dt, \label{A}
\end{align}
where $Q$ and $r_0$ are integration constants, $Q \in \mathbb{R}$ 
corresponds to the electric charge,  and 
$r_0 >0 $ corresponds to the radial position of the zero electric potential surface, 
which can (but not necessarily) be set equal to $+\infty$. 

Equation (\ref{Einstein}) has only three nontrivial components, i.e., the ($tt$), ($rr$), 
and ($\psi \psi$) components. The ($tt$) and ($rr$) components together 
give rise to
\begin{align*}
3\, \left( {\frac {d}{dr}}\phi \left( r \right)  \right) ^{2}-\phi
 \left( r \right) {\frac {d^{2}}{d{r}^{2}}}\phi \left( r \right) 
=0.
\end{align*}
Solving this equation, we get
\begin{align}
    \phi(r)= \pm\frac{1}{\sqrt{k\,r+b}}, \label{phibk}
\end{align}
even without providing a concrete form for the scalar potential $V(\phi)$. For 
$\phi(r)$ not to be singular at finite nonzero $r$, we require $k\ge 0,\,b\ge 0$, 
and $k$ and $b$ cannot be simultaneously zero. Note that the special choice $b\ne0$,
$k=0$ corresponds to constant $\phi$.

In this paper, we are interested in solutions with a nonconstant scalar field, so
we will be considering only the $k\ne 0$ branch of solutions. Inserting 
Eq. (\ref{phibk}) into the remaining field equations, Eqs. (\ref{Einstein}) and 
(\ref{scalar}), we can obtain explicit solutions for $f(r)$ and $V(\phi(r))$ as 
a function of $r$. However, the result is much too complicated. In particular, 
$f(r)$ contains terms that are proportional to the product of two logarithm 
functions and 
terms proportional to the special function 
$\mathrm{dilog}(r)$ defined as
\begin{align*}
 \mathrm{dilog}(x)=\int_1^{x}\frac{\ln(t)}{1-t}\mathrm{d} t.
\end{align*}
It does not make sense to reproduce the complicated result here. 
Significant simplifications arise if we take the choice
$k=\frac{1}{8B}$ and $b=\frac{1}{8}$. In this case, the scalar field becomes
\begin{align}
    \phi(r)=\pm \sqrt{\frac{8B}{r+B}}, \label{phir}
\end{align}
and the metric function reads
\begin{align}
  f(r)&=\left(3\,\beta-\frac{{Q}^{2}}{4}\,\right)
  + \left(2\beta-\frac{Q^2}{9}\right)\frac{B}{r}
  -Q^2\left(\frac{1}{2}+\frac{B}{3r} \right) \ln(r) 
  +\frac{{r}^{2}}{\ell^2}, \label{fr}
\end{align}
where $\beta$ and $\ell$ are integration constants. In order that the above $f(r)$ is a
solution, $V(\phi(r))$ as a function of $r$ must take a very special form.
We can invert $\phi(r)$ for $r$ and insert the result in $V(\phi(r))$ to get the scalar 
potential $V(\phi)$:
\begin{align}
V(\phi)&=-\frac{1}{\ell^2}+\frac{1}{512}\left(\frac {1}{\ell^2}
 +\frac{\beta}{B^2}\right)\phi^6 
 - \frac {1}{18432}\, \left(\frac{Q^2}{B^2}\right)\left(192\,{\phi}^{2}+
48{\phi}^{4}+5\,{\phi}^{6} \right) 
 \nonumber\\
 &
 +\frac{1}{3}\left(\frac{Q^2}{B^2}\right)\left[
\frac {2\phi^2}{ \left( 8-{\phi}^{2} \right)^2}
-\frac{1}{1024}{\phi}^{6}\ln 
 \left( {\frac {B \left( 8-{\phi}^{2}\right) }{{\phi}^{2}}} \right) \right].
 \label{Vr}
\end{align} 

The set of equations (\ref{A})--(\ref{Vr}) constitute a 
full set of exact solutions to the system defined by the action (\ref{action}), 
which has not been seen in the literature before. The
metric contains four parameters $B, \beta, \ell$, and $Q$. Among these, 
$B$ and $Q$ have already appeared in the solution for the scalar and Maxwell 
fields, respectively. 
$\Lambda\equiv -\frac{1}{\ell^2}$ appears in $V(\phi)$ as a constant term, 
which plays the role of a (bare) cosmological constant. In principle, the constant 
$\Lambda$ can either be positive, zero or negative.  However, if we wish to 
interpret the solution as a black hole solution, $\Lambda$ will
be necessarily negative, because in $(2+1)$ dimensions, smooth black hole 
horizons can exist only in the presence of a negative cosmological 
constant \cite{Ida}. That is why we 
adopted the notation $\Lambda = -\frac{1}{\ell^2}$ 
from the very beginning. The last parameter $\beta$ is related to the black hole mass 
$M$ via
\begin{align}
\beta=\frac{1}{3}\left(\frac{Q^2}{4}-M\right), \label{M}
\end{align} 
as will become clear in the degenerate case of a charged BTZ black hole 
\cite{Banados:1992wn} which corresponds to the case of $B=0$. At this point, we 
do not seem to have any principle to determine the allowed range for $\beta$. However,
the forthcoming physical analysis will make it clear that the $\beta$ value has to be
subjected to some constraints; otherwise, the solution will become physically 
unacceptable.

\subsection{Scalar potential}

Since $\lim_{\phi\to 0}V(\phi) =-\frac{1}{\ell^2}$, we may split $V(\phi)$ into
a sum
\[
V(\phi) =-\frac{1}{\ell^2} + U(\phi),
\]
where $U(\phi)$ encodes the true self-interaction of the scalar field $\phi$. 
Apart from the bare cosmological constant term and the following $\phi^6$ term, 
all the other terms are proportional to  $\frac{Q^2}{B^2}$, which implies that 
the scalar self-interaction is in a subtle balance with the Coulomb charge, 
though $\phi$ does not couple directly with the Maxwell field. The
seemingly complicated form of the potential ensures that
as $\phi\to 0$, the leading term in the power series expansion of $U(\phi)$ behaves
as $O(\phi^6)$, i.e.,
\begin{align}
U(\phi) \simeq \frac{1}{512}\left\{\frac{1}{\ell^2}+\frac{\beta}{B^2}
+\frac{1}{9}\left(\frac{Q^2}{B^2}\right)\left[
1-\frac{3}{2}\ln\left(\frac{8B}{\phi^2}\right)\right]
\right\}\phi^6 + \left(\frac{Q^2}{B^2}\right)\,O(\phi^8),
\label{uexp}
\end{align}
where only even powers of $\phi$ are present and the coefficients of all the 
$O(\phi^8)$ terms are positive. Some discussions are due 
here:
\begin{itemize}
\item If $Q=0$, then $U(\phi)$ degenerates into a $\phi^6$ potential, with coefficient
$\frac{1}{512}\left(\frac {1}{\ell^2} +\frac{\beta}{B^2}\right)$. If, in addition, 
$\beta=-\frac{B^2}{\ell^2}$, then the self-coupling of the scalar field 
vanishes. If $\beta<-\frac{B^2}{\ell^2}$, the potential has
a single extremum at $\phi=0$ which is a maximum, implying that 
the scalar potential is unbounded from below and the system is 
unstable under small perturbations in $\phi$. If $\beta>-\frac{B^2}{\ell^2}$, the 
single extremum becomes a minimum, which is stable against small perturbations in 
$\phi$. Thus, stability against small perturbations in $\phi$ requires 
$\beta>-\frac{B^2}{\ell^2}$.
\item If $Q\ne 0$, $U(\phi)$ will possess more than one extremum. This is more easily 
seen in the expanded form (\ref{uexp}). It is obvious that $\phi=0$ remains an 
extremum when $Q\ne 0$. Moreover, the term proportional to 
$\phi^{6}\ln(\phi^{2})$ possesses
two minima at some small nonzero $\phi$. The inclusion of the power series terms
may change the location of these minima, but the qualitative behavior of $U(\phi)$ 
remains unchanged, i.e., it has two minima at $\phi=\pm\phi_{\rm{min}} \ne 0$ 
and one maximum at $\phi=0$. 
\item From either the original potential (\ref{Vr}) or its expanded form 
(\ref{uexp}), it seems that we cannot take $B=0$. However, this observation is 
completely superficial because the scalar field $\phi$ also depends on $B$. If 
we substitute the value of $\phi(r)$ into $V(\phi)$ and then look at the resulting 
expression, it will be clear that the potential is perfectly regular at $B=0$.
\end{itemize}

\begin{figure}[h]
\centering
\begin{minipage}{.45\textwidth}
\centering
\includegraphics[width=\linewidth]{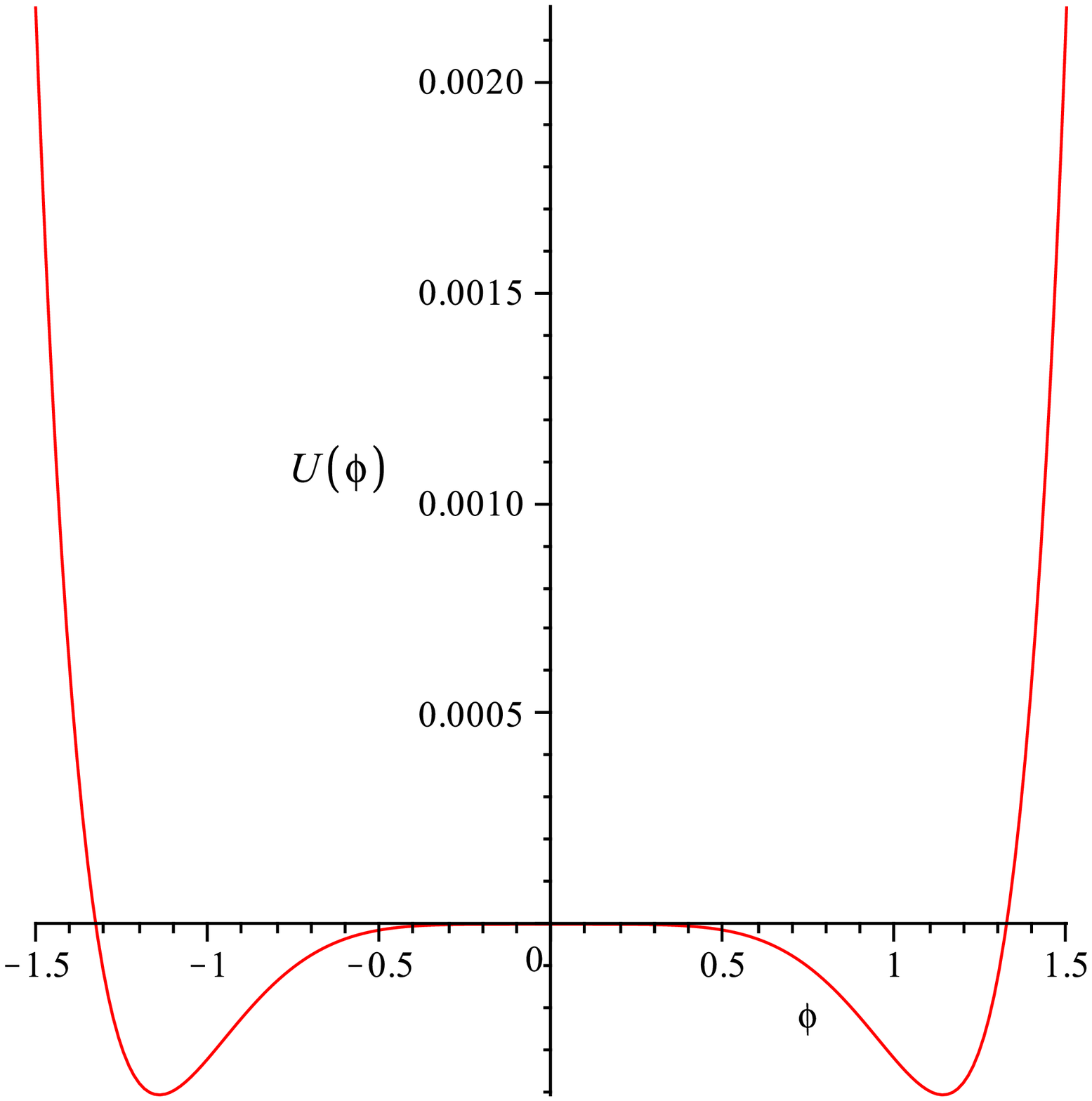}
\captionof{figure}{Plot of $U(\phi)$ vs $\phi$, with $B=1$, $\ell=1$, $\beta=-1$ 
and $Q=1$.} 
\label{FigV(phi)}
\end{minipage}\hspace{15pt}
\begin{minipage}{.45\textwidth}
\centering
\includegraphics[width=\linewidth]{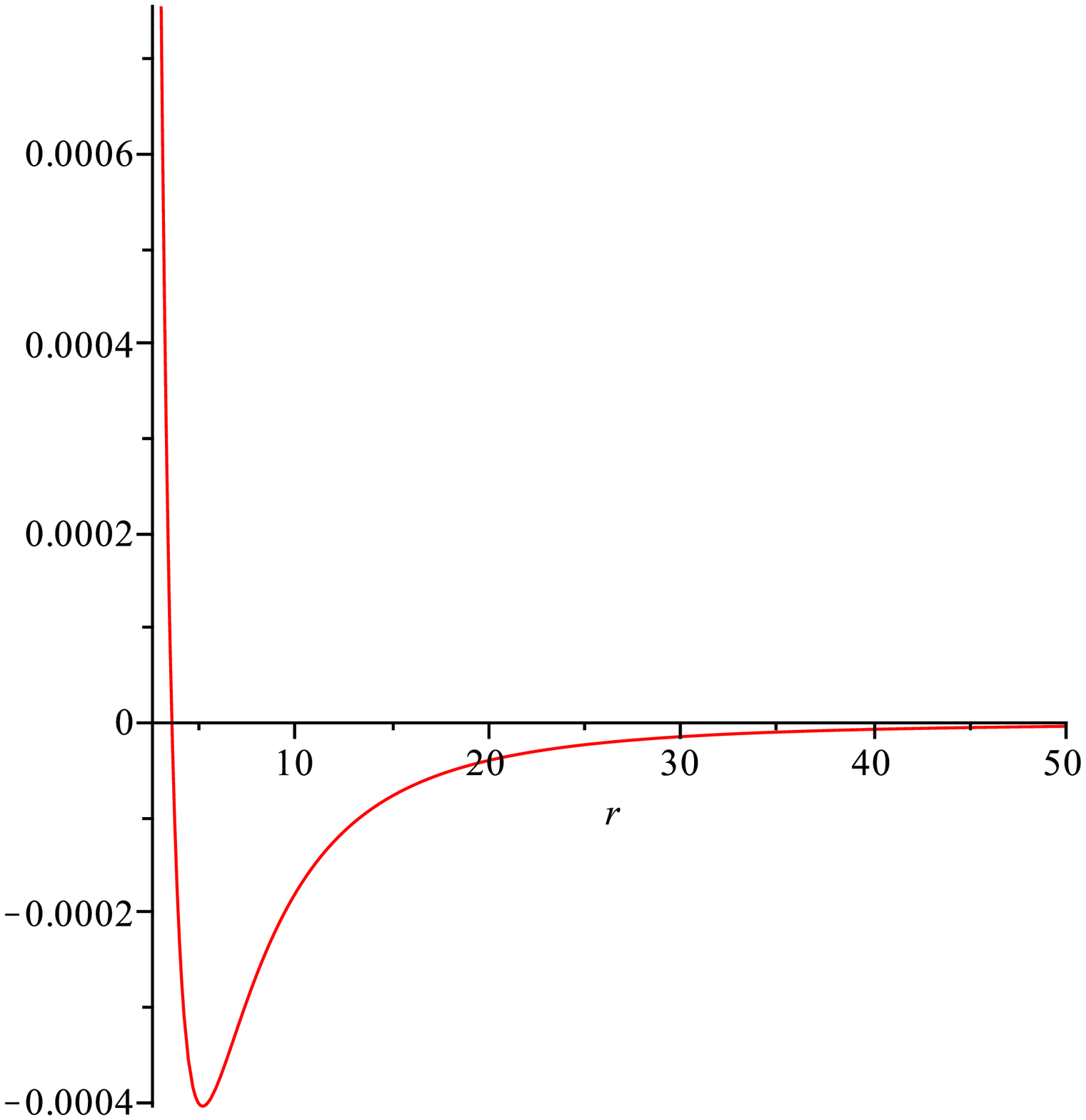}
\captionof{figure}{Plot of $U(\phi(r))$ vs $r$, 
with $B=1$, $\ell=1$, $\beta=-1$ and $Q=1$.} 
\label{FigV(phi(r))}
\end{minipage}
\end{figure}

In order to have more intuitive feelings about the scalar potential at $Q\ne 0$, we 
present a plot of $U(\phi)$ as a function of $\phi$ as well as a function of $r$. 
These are given in
Figs.\ref{FigV(phi)} and \ref{FigV(phi(r))} respectively.
If $\phi$ takes its value at the local maxima, i.e. $\phi=0$, the scalar field 
equation is automatically satisfied, and the scalar potential is exactly zero. 
The corresponding solution is an Einstein-Maxwell-AdS black hole. One may tends to 
think that if $\phi$ takes its value at any of the minima $\phi=\pm\phi_{\rm{min}}$, 
then the corresponding solution would correspond to the true vacuum of the 
system, with an effective cosmological constant 
$\Lambda_{\mathrm{eff}} =-\frac{1}{\ell^2}+U(\pm\phi_{\rm{min}})$ that is  
more negative than $-\frac{1}{\ell^2}$. However, this is not the case. If we take
$\phi=\pm\phi_{\rm{min}}\ne0$, then the field equation (\ref{scalar}) will force 
the Ricci scalar $R$ to be constant, which in turn requires $Q=0$. But when $Q=0$, 
the shape of the scalar potential $U(\phi)$ changes drastically, and the two 
minima at nonconstant $\phi$ totally disappear. Despite this, it is still 
an important observation that when $Q\ne 0$, the scalar potential $V(\phi)$ 
possesses two minima which are {\em smaller} than the cosmological constant 
$-\frac{1}{\ell^2}$. The physical explanation for these minima remains open. 
Because of the very complicated
form of the potential $U(\phi)$, we are unable to find the location of the two 
minima of $U(\phi)$ analytically. Nevertheless, it is easy to find the minima of
$U(\phi)$ numerically if the parameters $B,Q,\ell$, and $\beta$ were given numeric values. 
For instance, setting $B=Q=\ell=1, \beta=-1$, we find that the minima of
$U(\phi)$ are located at
\[
\phi=\pm\phi_{\rm{min}},\quad \phi_{\rm{min}} \simeq 1.139824 
\]
with the approximate value
\[
U(\pm\phi_{\rm{min}}) \simeq -0.000405
\]

\subsection{Some geometric properties of the solution}

To further characterize the geometry of the solution, we need to calculate some 
of the associated geometric quantities. First of all, the Ricci scalar
contains a curvature singularity at $r=0$ if $Q\ne 0$,
\begin{align*}
  R=-\frac {36\,{r}^{3}-3\,r{Q}^{2} \, {\ell}^{2}
  +2B\,{Q}^{2} \, {\ell}^{2}}{6{\ell}^{2}{r}^{3}}.
\end{align*}
Higher-order curvature invariants such as $R_{\mu\nu}R^{\mu\nu}$ and 
$R_{\mu\nu\rho\sigma}R^{\mu\nu\rho\sigma}$ are also singular at $r=0$, 
even if $Q=0$. However, the expressions for 
these invariants are much more complicated and unillustrative, so we do not 
reproduce them here. The Cotton tensor 
\[
C_{abc}=\nabla_c R_{ab}-\nabla_b R_{ac}+\frac{1}{4}\left(
\nabla_b R \, g_{ac}-\nabla_c R\, g_{ab}\right)
\]
is nonvanishing if either $B > 0$ or $Q\ne 0$,
\[
C_{trt}=-C_{ttr}=-\frac{1}{4}\,f \left( r \right) {\frac {d^{3}}{d{r}
^{3}}}f \left( r \right),
\]
\[
C_{\psi r \psi}=-C_{\psi\psi r}=-\frac{1}{4}\, 
\left( {\frac {d^{3}}{d{r}^{3}}}f
 \left( r \right)  \right) {r}^{2}.
\] 
In $(2+1)$ dimensions, the nonvanishing Cotton tensor signifies that the metric 
is nonconformally flat \cite{Garcia:2003p2522}. Thus the hairy ($B>0$) and 
charged ($Q\ne 0$) solutions are geometrically quite different from the case of 
the static uncharged BTZ black hole \cite{Banados:1992wn}.

\section{Special cases}

Before going into detailed analysis of our solution, we would like to point out some of 
the degenerated cases. Some of the degenerated cases have already been found in the 
literature.

\subsection{Charged BTZ black hole}
When $B=0$, the scalar field $\phi$ vanishes, and the system becomes 
the Einstein-Maxwell-AdS theory. The solution degenerates into the 
already known
static charged BTZ black hole \cite{Banados:1992wn}:
\begin{align*}
   f(r)&=-M-\frac{Q^2}{2}\ln(r)+\frac{r^2}{\ell^2},\\
  A_{\mu}dx^\mu&=-Q\ln\bigg(\frac{r}{r_0}\bigg)dt,\\
  V(\phi)&=-\frac{1}{\ell^2},\\
  \phi(r)&=0,
\end{align*}
where $M$ is the mass of the BTZ black hole. As mentioned earlier, 
although this solution may be stable in the Einstein-Maxwell-AdS theory, it is 
an unstable solution in the full theory (\ref{action}).

\subsection{Uncharged hairy AdS black hole}

The Maxwell field can be removed by setting $Q=0$. In this case we get an uncharged 
hairy AdS black hole solution
\begin{align*}
f(r)&=\left(3+\frac{2B}{r}\right)\beta+\frac{r^2}{\ell^2},\\
  A_{\mu}dx^\mu&=0,\\
  \phi(r)&=\pm\sqrt{\frac{8B}{r+B}},\\
  V(\phi)&=-\frac{1}{\ell^2}+\frac{1}{512}\left(\frac {1}{\ell^2}
 +\frac{\beta}{B^2}\right)\phi^6.
\end{align*}
Using Eq.(\ref{M}), we can change $\beta$ into $-\frac{M}{3}$ everywhere in the 
solution. In this case, the Ricci scalar becomes constant,
\begin{align*}
  R=-\frac{6}{\ell^2}=6\Lambda.
\end{align*}
However, higher-curvature invariants are still singular at $r=0$. This solution has already 
appeared in Sec. 5.1 of Ref.\cite{Nadalini:2007p2561}. As will become clear later, 
we need $\beta\in \left[-\frac{B^2}{\ell^2},0\right]$ in order for this 
solution to be physically well behaving.

\subsection{Conformally dressed black hole}
In addition to setting $Q=0$, we can choose $\beta=-\frac{B^2}{\ell^2}$ in the 
meantime. Under such conditions, we reproduce the conformally dressed 
black hole in $(2 + 1)$ dimensions \cite{Martinez:1996p2505}:
\begin{align*}
f(r)&=-\left(3+\frac{2B}{r}\right)\frac{B^2}{\ell^2}+\frac{r^2}{\ell^2},\\
  A_{\mu}dx^\mu&=0,\\
  \phi(r)&=\pm\sqrt{\frac{8B}{r+B}},\\
  V(\phi)&=-\frac{1}{\ell^2}.
\end{align*}
Note that althought the scalar field is still present, the self-interaction potential
$U(\phi)$ vanishes, thus making the scalar field a ``free'' massless field.

\subsection{Special charged hairy AdS black hole in three dimensions}

We may also choose $\beta=-\frac{B^2}{\ell^2}$ while keeping $Q$ nonvanishing. 
Then we get a special charged hairy AdS black hole
\begin{align*}
  f(r)&=-\left(\frac{3B^2}{\ell^2}+\frac{{Q}^{2}}{4}\,\right)
  -\left(\frac{2B^2}{\ell^2}+\frac{Q^2}{9}\right)\frac{B}{r}
  -Q^2\left(\frac{1}{2}+\frac{B}{3r} \right) \ln(r) 
  +\frac{{r}^{2}}{\ell^2}, \\
  \phi(r)&=\pm\sqrt{\frac{8B}{r+B}}, \\
  A_{\mu}dx^\mu&=-Q\ln\bigg(\frac{r}{r_0}\bigg)dt,\\
  V(\phi)&=-\frac{1}{\ell^2}
   - \frac {1}{18432}\, \left(\frac{Q^2}{B^2}\right)\left(192\,{\phi}^{2}+
   48{\phi}^{4}+5\,{\phi}^{6} \right) 
 \nonumber\\
 &
 +\frac{1}{3}\left(\frac{Q^2}{B^2}\right)\left[
\frac {2\phi^2}{ \left( 8-{\phi}^{2} \right)^2}
-\frac{1}{1024}{\phi}^{6}\ln 
 \left( {\frac {B \left( 8-{\phi}^{2}\right) }{{\phi}^{2}}} \right) \right].
\end{align*} 
The merit of this special case is that the scalar potential $U(\phi)$ 
is not fine-tuned with the cosmological constant.

\section{Horizon structures}

Among the three fields $g_{\mu\nu}$, $A_\mu$, and $\phi$, the latter two are easily 
understandable. $A_\mu$ is just the standard Coulomb potential in $(2+1)$ dimensions,
$\phi$ is a radially distributed scalar, which takes its maximum $\phi_{\mathrm{max}}
=\sqrt{8}$ at $r=0$ and decreases gradually to zero as $r\to +\infty$ when $B>0$. 

The metric $g_{\mu\nu}$ is more involved to be able to understand. 
This is because of the 
complicated form (\ref{fr}) of the metric function $f(r)$. As usual, the zeros of $f(r)$ 
(if any) will correspond to horizons in the metric. So, we need to find 
(at least the condition for the existence of) the zeros of $f(r)$.

\subsection{$Q=0$}

When $Q=0$, the functions $f(r)$ and $f'(r)$ are simplified drastically,
\begin{align*}
f(r)&=\beta\left(3 + \frac{2B}{r}\right)+\frac{{r}^{2}}{\ell^2},\\
f'(r)&=-\frac{2B\beta}{r^2}+\frac{2r}{\ell^2}.
\end{align*}
We may divide the solution in three subcases:
\begin{itemize}
\item If $\beta>0$, $f(r)$ will always be positive, implying that the metric 
corresponds to an asymptotically AdS${}_3$ spacetime containing a naked singularity. 
This is a physically uninteresting case.
\item If $\beta=0$, then the metric degenerates into the $(2+1)$-dimensional 
empty topological AdS spacetime. 
\item If, instead, $\beta<0$, then $f'(r)$ will remain positive for 
$r\in[0,+\infty)$. 
Meanwhile, $f(r)\to -\infty$ as $r\to 0$, $f(r)\to +\infty$ as $r\to +\infty$. 
This implies
$f(r)$ increases monotonically and, hence, contains exactly one zero. 
The zero of
$f(r)$ corresponds to the event horizon of a neutral AdS black hole with a scalar 
hair.
\end{itemize}

Combining with the analysis made in Sec. 2.2, we see that the physically 
acceptable range for the parameter $\beta$ is $\beta\in \left[-\frac{B^2}{\ell^2},0
\right]$ at $Q=0$.

\subsection{$Q\ne 0$}

The form of $f(r)$ with nonvanishing $Q$ is much more complicated than the 
$Q=0$ case. To find whether $f(r)$ has some zeros, we need to know the 
asymptotic behavior and the number of the extrema of $f(r)$.

From Eq.(\ref{fr}) it can be seen that $f(r)$ is dominated by the term 
$\frac{r^2}{\ell^2}$ in the far region, so, $f(r)\to +\infty$ as $r\to+\infty$.
On the other hand, as $r\to 0$, $f(r)$ is dominated by the term 
$-\frac{Q^2B}{3}\frac{\ln(r)}{r}$ if $B> 0$, or by the  
term $-\frac{Q^2}{2}\ln(r)$ if $B=0$. So, $f(r)$ always approaches $+\infty$ 
as $r\to 0$. Remember that we have excluded the possibility of choosing $B<0$
in order that $\phi(r)$ is not singular at finite nonzero $r$.
Combining the asymptotic behaviors at both ends, we see that
for $Q\ne 0$, $f(r)$ will approach $+\infty$ at both ends.
Therefore, $f(r)$ has to have some extrema, and the total number of 
extrema must be odd. Let us denote the location of the extrema of $f(r)$ by 
$r_{\mathrm{X}}$. If $f(r)$ has more than one extremum, 
an extra index may be adopted to distinguish these different extrema
if necessary. Clearly, the position $r_{\mathrm{X}}$ of every extremum 
of $f(r)$ obeys $f'(r_{\mathrm{X}})=0$, or, more conveniently
\begin{align}
(r_{\mathrm{X}})^2 f'(r_{\mathrm{X}}) = 0, \label{extrema}
\end{align}
where $r^2 f'(r)$ is given by the following expression
\begin{align}
r^2 f'(r) =2B \left(\frac{B^2}{9}+\beta\right) + \frac{1}{3}Q^2 B\ln(r) -
\frac{1}{2}Q^2 r+\frac{2r^3}{\ell^2}. \label{cd}
\end{align}
However, the converse needs not to be true: If $r^2 f'(r)$ happens to be zero at 
some of its extrema $r_i$ (not to be confused with $r_{\mathrm{X}}$), then 
$r_i$ will correspond to an inflection point of $f(r)$, rather than an extremum.

In order to determine the number of roots for $f(r)$, we need to 
consider two distinct cases, i.e. $B=0$ and $B>0$. For $B=0$, it is easy to 
get the location of the extremum of $f(r)$ using Eqs.(\ref{extrema}) and (\ref{cd}). 
The only real positive extremum of $f(r)$ in this case is located at 
$r=r_{\mathrm{X}}=
\frac{1}{2}|Q|\ell$. Inserting the value of $r_{\mathrm{X}}$ and $B=0$ into the
solution (\ref{fr}), we get the following:
\begin{itemize}
\item If $\beta<\frac{Q^2}{6}\ln(r_{\mathrm{X}})$, we have $f(r_{\mathrm{X}})<0$.
So $f(r)$ will have two zeros, each corresponding to a black hole horizon. Among 
these, the outer horizon is the event horizon. This case corresponds to a charged
nonextremal AdS black hole without the scalar hair.
\item If $\beta=\frac{Q^2}{6}\ln(r_{\mathrm{X}})$, we have $f(r_{\mathrm{X}})
=0$. It is evident that $r_{\mathrm{ex}}$ is the only root of $f(r)$, which 
corresponds to the horizon of a charged extremal AdS black hole without the scalar 
hair. In this particular case, it may be better to replace $r_{\mathrm{X}}$ with
$r_{\mathrm{ex}}$, which stands for the radius of the extremal black hole.
\item If $\beta>\frac{Q^2}{6}\ln(r_{\mathrm{X}})$, then $f(r_{\mathrm{X}})
>0$, there will be no zeros for the function $f(r)$, so the metric becomes 
an asymptotically AdS spacetime with a naked singularity at the origin, which is a 
physically uninteresting case.
\end{itemize}

The case with $B>0$ is much more complicated compared to the $B=0$ case. It is 
impossible to solve (\ref{extrema}) analytically to get $r_{\mathrm{ex}}$,
so we turn to look at the extrema of $r^2 f'(r)$. We introduce the following 
polynomial function:
\begin{align}
g(r)\equiv 6\ell^2\, r\, \frac{d}{dr}\left[r^2 f'(r) \right]
= 36r^3 -3Q^2 \ell^2 r +2B Q^2 \ell^2. \label{gri}
\end{align}
Every real positive root of $g(r)$ will correspond to an 
extremum or an inflection point of 
$r^2 f'(r)$, and the collection of signs of $r^2 f'(r)$ at its extrema will 
determine the number of roots thereof. Fortunately, the function $g(r)$ is 
simple enough so that its roots can be found analytically. In the appendix, 
we shall present the details about the roots of $g(r)$.

According to the appendix, the number of real positive roots of $g(r)$ will change 
when the value of the parameter $B$ crosses $\frac{|Q|\ell}{6}$. The significance 
of this change in the number of real positive roots will be best illustrated if we 
look at the extremum of $g(r)$.
Taking the first derivative of $g(r)$ with respect
to $r$ and finding the root of the resulting expression, we find that $g(r)$ has 
only one real positive extremum located at $r=\frac{|Q|\ell}{6}$. 
Clearly this is a minimum. Substituting this value of $r$ into $g(r)$ itself, 
we find the minimum value of $g(r)$, which reads
\begin{align}
g_{\mathrm{min}}=2B Q^2\ell^2 -\frac{1}{3}Q^3 \ell^3.
\end{align}
If $B>\frac{|Q|\ell}{6}$, then $g_{\mathrm{min}}$ is positive, which implies that
$g(r)$ has no real positive root, which in turn implies that $r^2 f'(r)$ 
has no extrema for $r>0$, so that $f'(r)$ has only one root, i.e. $f(r)$ has 
only one extremum.
If $B=\frac{|Q|\ell}{6}$, then $g(r)$ is zero at its extremum. This means that 
$g(r)$ has only one root that is located at its minimum.  This implies 
that the minimum of $g(r)$ corresponds to an inflection point 
rather than an extremum of $r^2f'(r)$. So, in
the end, $r^2f'(r)$ still has no extremum for $r>0$, resulting in the 
conclusion that $f(r)$ has only a single extremum for $r>0$.

The problem becomes more complicated when $0<B<\frac{|Q|\ell}{6}$. In this case,
$g(r)$ has three roots, two of which are positive. Therefore, $r^2 f'(r)$
will also have two extrema for $r>0$. Among these, the extremum at $r_1$ is a 
minimum, and that at $r_2$ is a maximum, and we have $r_2<r_1$.

Now let us assume $0<B<\frac{|Q|\ell}{6}$. Since $r_1$ and $r_2$ are both real 
positive roots of $g(r)$, we have 
\begin{align*}
r_i^3 = \frac{Q^2 \ell^2 r_i}{12}  - \frac{B Q^2 \ell^2}{18},\qquad i=1,2.
\end{align*}
Inserting this into Eq.(\ref{cd}), we get
\begin{align}
r_i^2 f'(r_i)&=2B\bigg(w(r_i)-\beta\bigg), \label{rfex}
\end{align}
where
\begin{align*}
     w(r_i)&\equiv\frac{Q^2}{6B}\left(B\ln(r_i)-r_i-B\right).
\end{align*}
Equation (\ref{rfex}) gives the value that $r^2 f'(r)$ must take at any of its extrema. 
In particular, the analysis made in the appendix implies $w(r_1)<w(r_2)$.

Depending on the value of the parameter $\beta$, $r_i^2 f'(r_i)$ will take different 
signs:
\begin{itemize}
\item If $\beta=w(r_1)$ or $\beta=w(r_2)$, we have $r_1^2 f'(r_1)=0$ or 
$r_2^2 f'(r_2)=0$, i.e., $r^2 f'(r)$ is zero at one of its extrema. Besides this
accidental root, $r^2 f'(r)$ has another root that is not at its extrema. 
So, totally $r^2 f'(r)$ will have two roots, one of these (the normal root) 
corresponds to the extremum of $f(r)$, and the other (the accidental root at the 
extremum) corresponds to an inflection point of $f(r)$. So, in the end, $f(r)$ will 
have only one extremum in this case.

\item If $\beta<w(r_1)$ or $\beta>w(r_2)$, we have either $r_2^2 f'(r_2)>
r_1^2 f'(r_1)>0$ or $0>r_2^2 f'(r_2)>r_1^2 f'(r_1)$. In both cases, all the 
extrema of $r^2 f'(r)$ have the same signs, so there is only one positive 
root of $r^2 f'(r)$. Consequently, $f(r)$ will have only a single extremum for 
$r>0$.

\item If $w(r_1)<\beta<w(r_2)$, we find $r_2^2 f'(r_2)>0>r_1^2 f'(r_1)$. 
The two positive extrema of $r^2 f'(r)$ have different signs, indicating that 
the curve for $r^2 f'(r)$ will cross the horizontal axes three times. Therefore,
there are three positive roots for $r^2 f'(r)$, each corresponding to an 
extremum of $f(r)$.
\end{itemize}

Summarizing the above discussions, we make the following conclusion on the number of
extrema for the metric function $f(r)$:
\begin{itemize}
\item If $B=0$ or $B\ge \frac{|Q|\ell}{6}$, $f(r)$ has only a single extremum;
\item If $0<B< \frac{|Q|\ell}{6}$, the number of extrema for $f(r)$ depends on the 
value of the parameter $\beta$. Explicitly,
\begin{itemize} 
\item if $\beta \le w(r_1)$ or $\beta \ge w(r_2)$, $f(r)$ still has only a single 
extremum;
\item if $w(r_1)<\beta<w(r_2)$ will have three extrema.
\end{itemize}
Therefore, the horizon structure of our solution at $Q\ne 0$ will depend crucially 
on the range of parameters.
\end{itemize}

Since at any extremum $r_{\mathrm{X}}$ of $f(r)$ we have 
$f'(r_{\mathrm{X}})=0$, 
an arbitrary multiple of $f'(r_{\mathrm{X}})$ 
can be added to $f(r_{\mathrm{X}})$ to yield a simplified 
expression for $f(r_{\mathrm{X}})$. Specifically, we take the 
following combination:
\begin{align}
  p(r_{\mathrm{X}})&\equiv \frac{B}{r_{\mathrm{X}}+B}
  \bigg(f(r_{\mathrm{X}})+f'(r_{\mathrm{X}})
\frac{r_{\mathrm{X}}(9r_{\mathrm{X}}+6B)}{6B}\bigg)\nonumber\\
&=\frac{B}{r_{\mathrm{X}}+B}f(r_{\mathrm{X}}) =36\,(r_{\mathrm{X}})^{3}
-9\,{Q}^{2}{\ell}^{2}\,r_{\mathrm{X}}-4\,B{Q}^{2}{\ell}^{2}. \label{px}
\end{align}
Clearly, $p(r_{\mathrm{X}})$ and $f(r_{\mathrm{X}})$ always have the same sign, 
so the collection of signs of $p(r_{\mathrm{X}})$ at all the extrema 
$r_{\mathrm{X}}$ of $f(r)$ will determine the number of roots of $f(r)$.

Let us consider the ``extremal case'' defined via 
$f(r_{\mathrm{ex}})=p(r_{\mathrm{ex}})=0$ and $f'(r_{\mathrm{ex}})=0$. In this case,
$f(r)$ happens to be zero at one of its extrema. Since the zeros of 
$f(r_{\mathrm{ex}})$ and $p(r_{\mathrm{ex}})$ always coincide, we can try to find 
the zeros of $p(r_{\mathrm{ex}})$ to get the value 
of $r_{\mathrm{ex}}$. 

Since we have $B\ge 0$, we may assume
\begin{align}
  B&=\frac{a}{4}|Q|\ell,\qquad
  r_{\mathrm{ex}}=\rho  |Q|\ell, \label{Brex}
\end{align}
so that the equation $p(r_{\mathrm{ex}})=0$ becomes
\begin{align}
  36 \rho ^3-9 \rho -a =0. \label{a-eq}
\end{align}
The condition $a\ge0$ implies $\rho\ge \frac{1}{2}$, where the lower bound for 
$\rho$ corresponds to 
$a=0$, i.e. $B=0$, as discussed previously. Generically, eq.(\ref{a-eq}) can 
have three zeros. However, only one of these is real positive and 
lies in the range $\rho\in \left[\frac{1}{2}, +\infty\right)$, 
which is given in terms of $a$ as
\begin{align}
\rho = \frac{z}{6}+\frac{1}{2z},
\quad z=\sqrt [3]{3\,a+3\,\sqrt {{a}^{2}-3}}, \quad a\ge 0. \label{prho}
\end{align}
The detailed analysis on the solution to (\ref{a-eq}) can be carried out in exactly 
the same way as is done in the appendix for the similar equation (\ref{greq0}). 

Now, substituting Eqs. (\ref{M}) and (\ref{Brex}) into the equation $f'(r_{\mathrm{ex}})=0$, 
we get, 
\begin{align*}
 \beta=\frac{Q^2}{6}\ln(r_{\mathrm{ex}})
  =\frac{Q^2}{6}\ln\left(\rho |Q|\ell\right).
\end{align*}
This is the same condition that appeared in the $B=0$ case.
The only difference lies in the fact that $\rho$ is fixed at the value $\frac{1}{2}$
when $B=0$, while for $B\ne 0$,  its value is given by Eq.(\ref{prho}). 

We have already made it clear that when $B=0$ or $B\ge \frac{|Q|\ell}{6}$, or
when $0<B<\frac{|Q|\ell}{6}$ with $\beta \le w(r_1)$ or $\beta \ge w(r_2)$, $f(r)$ 
always has only a single minimum. So, for these parameter ranges, the 
zero $r_{\mathrm{ex}}$ of $f(r)$ as described in Eqs.(\ref{Brex}) and (\ref{prho}) is
the only zero and minimum of $f(r)$. So, these cases correspond to 
extremal black holes with horizon radius 
\[
r_{\mathrm{ex}}=\exp\left(\frac{6\beta}{Q^2}\right).
\]

If at the extremum 
$r_{\mathrm{X}}$ of $f(r)$, $p(r_{\mathrm{X}})$ fails to be zero, then the 
corresponding solution will not correspond to an extremal black hole. Let us now 
consider such cases in more detail. 

For all $r_{\mathrm{X}}\ge r_{\mathrm{ex}}=\rho|Q|\ell$, we have
\[
p'(r_{\mathrm{X}}) = 9(12r_{\mathrm{X}}^2 - Q^2\ell^2) \ge p'(r_{\mathrm{ex}})
= 9\left(3 Q^2\ell^2- Q^2\ell^2\right) =18 Q^2 \ell^2>0,
\]
i.e. $p(r_{\mathrm{X}})$ increases monotonically for 
$r_{\mathrm{X}}\ge r_{\mathrm{ex}}$. So, if $p(r_{\mathrm{X}})<0$, then  
$r_{\mathrm{X}}$ must be located 
to the left of $r_{\mathrm{ex}}$, i.e. $r_{\mathrm{X}}< r_{\mathrm{ex}}$.
However, if $p(r_{\mathrm{X}})>0$, we cannot deduce from above that 
$r_{\mathrm{X}}> r_{\mathrm{ex}}$.

Consider the following identity:
\begin{align*}
  &\bigg(1-\frac{B}{r_{\mathrm{X}}+B}\bigg)f(r_{\mathrm{X}})
  =3\beta-\frac{{Q}^{2}}{2}\,\ln  \left( r_{\mathrm{X}} \right).
\end{align*}
This is equivalent to
\begin{align}
\beta=\frac{{Q}^{2}}{6}\,\ln  \left( r_{\mathrm{X}} \right)
+\frac{r_{\mathrm{X}}}{3(r_{\mathrm{X}}+B)}f(r_{\mathrm{X}}). \label{betafr}
\end{align}
If $f(r_{\mathrm{X}})<0$, then
\begin{align}
\beta<\frac{Q^2}{6}\ln\left(r_{\mathrm{X}}\right)
<\frac{Q^2}{6}\ln\left(r_{\mathrm{ex}}\right).  \label{non-extr}
\end{align}
In Eq.(\ref{non-extr}), $r_{\mathrm{X}}$ may be extremely close to 
$r_{\mathrm{ex}}$, while still keeping $r_{\mathrm{X}}<r_{\mathrm{ex}}$. So, we may 
think of 
\begin{align}
\beta<\frac{Q^2}{6}\ln\left(r_{\mathrm{ex}}\right)  \label{non-extr2}
\end{align}
to be the condition that must be imposed on the parameter $\beta$ in order for the 
metric to have two disjoint horizons, to behave as a nonextremal black hole. 
If $f(r_{\mathrm{X}})>0$, then 
\begin{align}
\beta>\frac{Q^2}{6}\ln\left(r_{\mathrm{X}}\right). \label{naked}
\end{align}
Under this condition, the metric contains no horizons and corresponds to 
an asymptotically AdS${}_3$ spacetime with a naked 
singularity. 

What remains untouched is the real troublesome case with $0<B<\frac{|Q|\ell}{6}$
and $w(r_1)<\beta<w(r_2)$. In this parameter range, we 
have to be careful about how many zeros there are for $f(r)$, because $f(r)$ has 
three extrema. Let us denote the location of the three extrema by $r_{\mathrm{X1}},
r_{\mathrm{X2}}$ and $r_{\mathrm{X3}}$ respectively. Let  $r_{\mathrm{X1}},
r_{\mathrm{X2}}$ and $r_{\mathrm{X3}}$ be ordered such that $r_{\mathrm{X1}}<
r_{\mathrm{X2}}<r_{\mathrm{X3}}$. At these points, $r^2 f'(r)$ vanishes, 
and its two 
positive extrema locate between these zeros, i.e. 
\[
r_{\mathrm{X1}}< r_2 < r_{\mathrm{X2}}< r_1 < r_{\mathrm{X3}}.
\]
From the appendix, we have $\frac{|Q|\ell}{6}<r_1<\frac{\sqrt{3}|Q|\ell}{6}<
\frac{|Q|\ell}{2}$, so, 
\[
r_{\mathrm{X2}}<\frac{|Q|\ell}{2}.
\]
Therefore, according to (\ref{px}), we have
\begin{align*}
  p(r_{\mathrm{X2}})&=\frac{B}{r_{\mathrm{X2}}+B}f(r_{\mathrm{X2}}) 
  =36\,(r_{\mathrm{X2}})^{3}-9\,{Q}^{2}{\ell}^{2}\,r_{\mathrm{X2}}
  -4\,B{Q}^{2}{\ell}^{2}\\
  &=36r_{\mathrm{X2}}\left(r_{\mathrm{X2}}^2-\frac{Q^2\ell^2}{4}\right)
  -4\,B{Q}^{2}{\ell}^{2}<0,
\end{align*}
i.e. $f(r_{\mathrm{X2}})<0$. Since $r_{\mathrm{X2}}$ lies in between 
$r_{\mathrm{X1}}$ and $r_{\mathrm{X3}}$, it corresponds to the {\em local maximum}
of $f(r)$, so the above result implies that $f(r)$ is negative at all its three 
extrema. Combining with the asymptotic behavior, we deduce that $f(r)$ have 
precisely two zeros in this case, so, whenever $f(r)$ has three extrema, the solution 
corresponds to a nonextremal charged hairy black hole with AdS asymptotics.
Please note that, for this case, we have
\begin{align*}
  \beta&<w(r_2)=\frac{Q^2}{6B}\left(B\ln(r_2)-r_2-B\right)
  <\frac{Q^2}{6}\ln(r_2)\\
  &\quad<\frac{Q^2}{6}\ln\left(\frac{|Q|\ell}{6}\right)
  <\frac{Q^2}{6}\ln\left(\frac{|Q|\ell}{2}\right)
  \leq\frac{Q^2}{6}\ln\left(\rho |Q|\ell\right)
  =\frac{Q^2}{6}\ln\left(r_{\mathrm{ex}}\right).
\end{align*}

Putting all the cases together, we see that the physically acceptable upper 
bound for $\beta$ at $Q\ne 0$ is
\[
\beta \le \frac{Q^2}{6}\ln\left(r_{\mathrm{ex}}\right).
\]
Depending on the value of $r_{\mathrm{ex}}$ determined by Eqs.(\ref{Brex}) and
(\ref{prho}), this upper bound can be either positive or negative. Meanwhile, for 
$Q\ne 0$, the scalar potential $U(\phi)$ is dominated by the $O(\phi^8)$ terms 
when $\phi$ is big enough, so we do not need to worry about the existence 
of lower bound for the scalar potential. 
Accordingly, there is no lower bound for the parameter $\beta$.

\section{Effects of scalar and electric charges on the size of the black hole}

In this section, we shall restrict ourselves to the cases in which (a) 
black hole horizon(s) 
exist(s). The primitive goal of this section is to understand the effects of the scalar 
and electric charges on the size of the black hole. Of course, the best way to 
do this is to consider the effect of each charge independently.

\subsection{Effect of the scalar charge}

The parameter $B$ originates solely from the scalar field $\phi$, but it is also carried 
by the black hole solution, so this parameter may be regarded as a scalar charge of the 
hole. In order to understand the effect of the scalar charge $B$ on the size of the 
black hole, we need to separate the metric function $f(r)$ into $B$-independent and 
$B$-dependent parts. According to Eq.(\ref{fr}), $f(r)$ depends on $B$ only linearly,
so,
\begin{align*}
&f(r)= f(r)|_{B=0} + f_B(r),
\end{align*}
where
\begin{align}
&f_B(r)=B\,\frac{df(r)}{dB}=B\left[\left(2\beta-\frac{Q^2}{9}\right)\frac{1}{r}
  -\frac{Q^2}{3r} \ln(r)\right] . \label{fb}
\end{align}
Notice that the dominant term $\frac{r^2}{\ell^2}$ as $r\to+\infty$ is contained
in the $f(r)|_{B=0}$. Solving $\ln(r)$ out of the horizon condition $f(r)=0$ and 
substituting in (\ref{fb}), we get
\begin{align*}
&f_B(r)=\frac{B(Q^2\ell^2-6r^2)}{\ell^2(6B+9r)}, 
\end{align*}
where $r$ must be understood as the horizon radius. It is clear that $f_B(r)<0$ provided
$B>0$ and $r>\frac{|Q|\ell}{\sqrt{6}}$. We already know from the previous section 
that the horizon radius for the extremal black hole is $r_{\mathrm{ex}}=\rho |Q|\ell$
with $\rho\ge\frac{1}{2}>\frac{1}{\sqrt{6}}$, and, in general, the radius $r_+$ for the outer horizon of the nonextremal black hole is bigger yet than 
$r_{\mathrm{ex}}$; we 
see that the effect of inclusion of the $f_B(r)$ term in $f(r)$ is to make $f(r)$ more 
negative. Therefore, to compensate for this extra negative contribution, the 
$f(r)|_{B=0}$ term must become more positive in order to make a zero for $f(r)$.
In other words, the size of the black hole must increase as $B$ increases.

\subsection{Effect of the electric charge}

We can also separate $f(r)$ into its $Q^2$-independent and $Q^2$-dependent parts,
i.e.
\begin{align}
&f(r)= f(r)|_{Q=0} +  f_{Q^2}(r),
\end{align}
where
\begin{align}
& f_{Q^2}(r)=Q^2\,\frac{df(r)}{d(Q^2)}=Q^2\left[-\frac{1}{4}-\frac{B}{9r}
  -\left(\frac{1}{2}+\frac{B}{3r} \right) \ln(r)\right] .
\end{align}
The horizon condition is a proper balance between the $f(r)|_{Q=0}$  and the
$ f_{Q^2}(r)$ terms. In Sec. 4.1, it was shown that $f(r)|_{Q=0} \to -\infty$
as $r\to 0$ and increases monotonically with $r$. On the other hand, from the above
expression, we see that $ f_{Q^2}(r)\to +\infty$ as $r\to 0$ 
and {\em decreases monotonically} with $r$. 
Moreover, in the ``near end'' ($r\to 0$), the term $-\frac{1}{r}\ln(r)$ in  
$ f_{Q^2}(r)$ takes over the term $-\frac{1}{r}$ in  $f(r)|_{Q=0}$, and in the far
region ($r\to+\infty$), the term $\frac{r^2}{\ell^2}$ in $f(r)|_{Q=0}$ 
dominates, while $ f_{Q^2}(r)$ becomes increasingly negative. 
In effect, the inclusion of the $ f_{Q^2}(r)$ term in $f(r)$ results in two 
different consequences: In the near region, 
$f(r)$ develops a novel zero, which is the inner horizon for the charged black hole; 
in the far region, the original zero of $f(r)|_{Q=0}$ (now being the outer horizon 
of the charged black hole) gets increased with the inclusion of the 
$ f_{Q^2}(r)$ term, and as $Q^2$ increases, the radius of the outer horizon
also increases monotonically.

\section{Discussions}

In this paper, we obtained an exact solution for the Einstein-Maxwell-scalar theory 
in $(2+1)$ dimensions,
in which the scalar field couples to gravity nonminimally and also couples
to itself in a peculiar way. The solution is static, circularly symmetric, and the scalar self-potential is completely determined by staticness and the
circular symmetry of the 
solution. In particular, a negative cosmological constant naturally emerges 
as a constant term in the scalar potential if 
we require that, for certain ranges of parameters, the solution represents a 
charged hairy black hole. Under the proper choices of parameter values, our 
solution degenerates 
into some already known $(2+1)$-dimensional black hole solutions. 

When the electric charge $Q$ is nonzero, the scalar potential possesses three 
extrema, one maximum at $\phi=0$ and two minima at $\phi\ne0$. The scalar field
$\phi$ cannot stay at the constant value $\phi=\pm\phi_{\mathrm{min}}\ne0$; 
otherwise, the field equations will not be satisfied.

We also identified the conditions for the metric to behave as a charged extremal 
black hole, as an asymptotically AdS${}_3$ spacetime with a naked singularity at the 
origin, and as a charged nonextremal black hole. When black hole horizons exist, it 
is shown that the size of the (outer) horizon increases monotonically with both the 
scalar charge and the electric charge.

Some of the related properties and duality relations will become instantly 
interesting further tasks to be worked out. These are:
\begin{itemize}
\item the thermodynamic quantities and the associated laws of thermodynamics;
\item the properties of the boundary CFT, if any;
\item that these days the fluid dual of AdS gravity is an active area of study and it will 
be interesting to ask whether there is  fluid dual of the black hole solution given 
in this paper;
\item that the solution considered in this paper is only static and circularly symmetric, and it would be interesting to ask whether one can find more 
complicated solutions to 
the same theory (for instance, it is interesting to ask whether one can find 
rotationally symmetric solutions and determine the scalar self-interaction solely by 
the form of the metric ansatz);
\item that the scalar field in model of this paper is neutral and does not couple to the 
electromagnetic field, and it would also be interesting to allow the scalar field to 
become complex and thus couple directly to the electromagnetic field.
\end{itemize} 
We leave the answer to all these problems for future work.

\section*{Appendix: Roots of the function $g(r)$}

In this appendix we shall solve the root of the function
$g(r)$ given by Eq.(\ref{gri}), i.e., the root of the equation
\begin{align}
36r^3 -3Q^2 \ell^2 r +2B Q^2 \ell^2=0. \label{greq0}
\end{align}
With the aid of a computer algebra system like Maple, it is easy to find 
that the roots of the above equation are given by
\begin{align*}
r_1&=\frac{1}{6}\left(z+\frac{Q^2\ell^2}{z}\right),\\
r_2&=\frac{1}{12}\left[-z-\frac{Q^2\ell^2}{z}
   +i\sqrt{3}\left(z-\frac{Q^2\ell^2}{z}\right)\right],\\
r_3&=\frac{1}{12}\left[-z-\frac{Q^2\ell^2}{z}
   -i\sqrt{3}\left(z-\frac{Q^2\ell^2}{z}\right)\right],
\end{align*}
where
\begin{align}
z=\sqrt[3]{-6B Q^2 \ell^2 +\sqrt{36 B Q^4 \ell^4-Q^6\ell^6}}.
\label{z1}
\end{align}
Among the three roots of $g(r)$, only the real positive roots are relevant to 
our problem. So we need to identify which and how many of the roots are real 
positive. 

It is evident that if $B\ge\frac{|Q|\ell}{6}$, then $z$ is real.  It follows that
if $B> \frac{|Q|\ell}{6}$, then $r_1$ is real but negative, while both $r_2$ 
and $r_3$ are complex. If $B=\frac{|Q|\ell}{6}$, then $r_1$ and $r_2$ will 
become degenerate and both are real positive. In this case $r_3$ is negative. 
If $B<\frac{|Q|\ell}{6}$, 
$z$ becomes complex. In this case, it is not too difficult to see that
the modulus of $z$ is equal to $|Q|\ell$, so we can denote $z$ as
\begin{align*} 
z=|Q|\ell\,(\cos\theta + i\sin\theta).
\end{align*}
Comparing this expression with the original definition (\ref{z1}), we see that
$\theta$ must take value in the range
\begin{align*}
\frac{\pi}{6}<\theta<\frac{\pi}{3},
\end{align*}
and the three roots of $g(r)$ can be reparametrized as
\begin{align}
&r_1=\frac{|Q|\ell}{3}\cos(\theta),\\
&r_2=-\frac{|Q|\ell}{3}\cos\left(\theta+\frac{\pi}{3}\right),\\
&r_3=-\frac{|Q|\ell}{3}\cos\left(\theta-\frac{\pi}{3}\right).
\end{align}
It is easy to see that
\begin{align}
&r_1 \in \left(\frac{|Q|\ell}{6}, \frac{\sqrt{3}}{6}|Q|\ell \right),\\
&r_2 \in \left(0, \frac{|Q|\ell}{6}\right),\\
&r_3 \in \left(-\frac{\sqrt{3}}{6}|Q|\ell, -\frac{|Q|\ell}{6} \right).
\end{align}
All three roots are real; however, only $r_1$ and $r_2$ are positive.
Clearly, $r_1$ is the bigger root of $g(r)$.


\begin{thebibliography}{50}

\bibitem{JK}S. Deser, R. Jackiw and S. Templeton, ``Topological massive gauge 
  theories'', Ann. Phys. 140, 372 (1982)

\bibitem{JK2}S. Deser, R. Jackiw and S. Templeton, ``Three-dimensional 
  massive gauge theories'', Phys. Rev. Lett. 48, 975 (1982).

\bibitem{Jackiw}S.~Deser and R.~Jackiw, ``Three-dimensional cosmological 
  gravity: Dynamics of constant curvature,'' Ann. Phys. 153 (1984) 405-416.

\bibitem{Giddings}S. Giddings, J. Abbott, and K. Kuchar, Einstein's theory in a 
  three-dimensional space-time, Gen. Relat. Grav. 16, 751 (1984).

\bibitem{Banados:1992wn}
  M.~Ba\~nados, C.~Teitelboim and J.~Zanelli,
  ``The black hole in three dimensional spacetime,''
  Phys.\ Rev.\ Lett.\  {\bf 69}, 1849 (1992)  [\eprint{hep-th/9204099}].

\bibitem{Martinez:1999p2523}
C.~Martinez, C.~Teitelboim, and J.~Zanelli, ``Charged Rotating Black Hole in
  Three Spacetime Dimensions,'' Phys.Rev. D61 (2000) 104013 
  [\eprint{hep-th/9912259}].

\bibitem{Brown:1986p547}
J.~Brown and M.~Henneaux, ``Central charges in the canonical realization of
  asymptotic symmetries: an example from three dimensional gravity,'' 
  Commun. Math. Phys. {\bf 104} (1986), no.~2, 207--226.

\bibitem{Henneaux:2002p2538}
M.~Henneaux, C.~Martinez, R.~Troncoso, and J.~Zanelli, ``Black holes and
  asymptotics of 2+1 gravity coupled to a scalar field,'' Phys.Rev. D65 (2002) 104007 
  [\eprint{hep-th/0201170}].

\bibitem{Hasanpour:2011p2274}
M.~Hasanpour, F.~Loran, and H.~Razaghian, ``Gravity/CFT correspondence for
  three dimensional Einstein gravity with a conformal scalar field,'' 
  Nucl. Phys. B867 (2013) 483-505 [\eprint{1104.5142}]. 
  
\bibitem{Martinez:1996p2505}
  C.~Martinez and J.~Zanelli,
  ``Conformally dressed black hole in (2+1)-dimensions,''
  Phys.\ Rev.\ D {\bf 54}, 3830 (1996)  [\eprint{gr-qc/9604021}].

\bibitem{Natsuume:1999p2556}
M.~Natsuume, T.~Okamura, and M.~Sato, ``Three-Dimensional Gravity with
  Conformal Scalar and Asymptotic Virasoro Algebra,'' Phys.Rev. D61 (2000) 104005 
  [\eprint{hep-th/9910105}].
  
\bibitem{Correa:2010p2540}
F.~Correa, C.~Martinez, and R.~Troncoso, ``Scalar solitons and the microscopic
  entropy of hairy black holes in three dimensions,'' JHEP 1101:034,2011
   [\eprint{1010.1259}].
  
\bibitem{Correa:2011p2543}
F.~Correa, C.~Martinez, and R.~Troncoso, ``Hairy black hole entropy and the
  role of solitons in three dimensions,'' [\eprint{1112.6198}]. 

\bibitem{Correa:2012p2547}
F.~Correa, A.~Faundez, and C.~Martinez, ``Rotating hairy black hole and its
  microscopic entropy in three spacetime dimensions,'' [\eprint{1211.4878}].
  
\bibitem{Zeng:2009p2536}
D.~F. Zeng, ``An Exact Hairy Black Hole Solution for AdS/CFT
  Superconductors,'' [\eprint{0903.2620}].

\bibitem{Colgain:2010p108}
E.~{\'O}. Colg{\'a}in and H.~Samtleben, ``3D gauged supergravity from wrapped
  M5-branes with AdS/CMT applications,'' 	JHEP 1102:031,2011
   [\eprint{1012.2145}].

\bibitem{Chen:2013p2529}
B.~Chen, Z.~Xue, and J.~ju~Zhang, ``Note on Thermodynamic Method of Black
  Hole/CFT Correspondence,'' [\eprint{1301.0429}].

\bibitem{Blagojevic:2013p2533}
M.~Blagojevi{\'c}, B.~Cvetkovi{\'c}, O.~Miskovic, and R.~Olea, ``Holography in
  3D AdS gravity with torsion,'' [\eprint{1301.1237}].
  
\bibitem{Cai:2012vr} 
  R.~-G.~Cai, L.~Li, Z.~-Y.~Nie and Y.~-L.~Zhang,
  ``Holographic Forced Fluid Dynamics in Non-relativistic Limit,''  
  Nucl.\ Phys.\ B {\bf 864}, 260 (2012)  [\eprint{1202.4091}].  

\bibitem{Myung:2008p2548}
Y.~S. Myung, ``Phase transition for black holes with scalar hair and
  topological black holes,'' Phys.Lett.B663:111-117,2008 [\eprint{0801.2434}].

\bibitem{Eune:2013p2524}
M.~Eune and W.~Kim, ``Hawking-Page phase transition in BTZ black hole
  revisited,'' [\eprint{1301.0395}]. 

\bibitem{Kamata:1995p2516}
M.~Kamata and T.~Koikawa, ``The electrically charged BTZ black hole with
  self(anti-self) dual Maxwell field,'' Phys.Lett. B353 (1995) 196-200 
  [\eprint{hep-th/9505037}].

\bibitem{Martinez:2006p2553}
C.~Martinez and R.~Troncoso, ``Electrically charged black hole with scalar
  hair,'' Phys.Rev.D74:064007,2006 [\eprint{hep-th/0606130}].

\bibitem{Martinez:2004p2545}
C.~Martinez, R.~Troncoso, and J.~Zanelli, ``Exact black hole solution with a
  minimally coupled scalar field,'' 	Phys.Rev. D70 (2004) 084035 
  [\eprint{hep-th/0406111}].

\bibitem{Bekenstein:1996p2554}
J.~D. Bekenstein, ``Black hole hair: twenty-five years after,'' 
  [\eprint{gr-qc/9605059}].

\bibitem{Chan:1996p2544}
K.~C.~K. Chan, ``Modifications of the BTZ black hole by a dilaton/scalar,''
  Phys.Rev. D55 (1997) 3564-3574 [\eprint{gr-qc/9603038}].

\bibitem{Abramo:2003p2564}
L.~R. Abramo, L.~Brenig, E.~Gunzig, and A.~Saa, ``A note on dualities in
  Einstein's gravity in the presence of a non-minimally coupled scalar field,''
  Mod.Phys.Lett. A18 (2003) 1043 [\eprint{gr-qc/0305009}].

\bibitem{Nadalini:2007p2561}
M.~Nadalini, L.~Vanzo, and S.~Zerbini, ``Thermodynamical properties of hairy
  black holes in n spacetimes dimensions,'' 
  Phys.Rev.D77:024047,2008 [\eprint{0710.2474}].

\bibitem{Acena:2012p2551}
A.~Acena, A.~Anabalon, and D.~Astefanesei, ``Exact hairy black brane solutions
  in AdS${}_{5}$ and holographic RG flows,'' [\eprint{1211.6126}]. 
  
\bibitem{Tafel:2011p2531}
J.~Tafel, ``Static spherically symmetric black holes with 
  scalar field,'' [\eprint{1112.2687}].

\bibitem{Pugliese:2013p2525}
D.~Pugliese and J.~A.~V. Kroon, ``On the evolution equations for a
  self-gravitating charged scalar field,'' [\eprint{1301.0461}].

\bibitem{Aparicio:2012p2461}
J.~Aparicio, D.~Grumiller, E.~Lopez, I.~Papadimitriou, and S.~Stricker,
  ``Bootstrapping gravity solutions,'' [\eprint{1212.3609}]. 
  
\bibitem{Dias:2011p376}
O.~J.~C. Dias, G.~T. Horowitz, and J.~E. Santos, ``Black holes with only one
  Killing field,'' JHEP 1107:115,2011 [\eprint{1105.4167}]. 
  
\bibitem{Cadoni:2011p911}
M.~Cadoni, S.~Mignemi, and M.~Serra, ``Exact solutions with AdS asymptotics of
  Einstein and Einstein-Maxwell gravity minimally coupled to a scalar field,''
  [\eprint{1107.5979}]. 
  
\bibitem{Kolyvaris:2009p2549}
T.~Kolyvaris, G.~Koutsoumbas, E.~Papantonopoulos, and G.~Siopsis, ``A New Class
  of Exact Hairy Black Hole Solutions,'' 
  Gen.Rel.Grav.43:163-180,2011 [\eprint{0911.1711}].

\bibitem{Degura:1998p2546}
Y.~Degura, K.~Sakamoto, and K.~Shiraishi, ``Black Holes with Scalar Hair in
  (2+1) dimensions,'' Grav.Cosmol. 7 (2001) 153-158 [\eprint{gr-qc/9805011}].

\bibitem{Banados:2005p2542}
M.~Ba\~nados and S.~Theisen, ``Scale Invariant Hairy Black Holes,'' 
 Phys.Rev. D72 (2005) 064019  [\eprint{hep-th/0506025}].

\bibitem{BerredoPeixoto:2002p2539}
G.~de~Berredo-Peixoto, ``On the static solutions in gravity with massive scalar
  field in three dimensions,'' Class. Quantum Grav. 20 (2003) 3983-3992 
  [\eprint{gr-qc/0208026}].

\bibitem{Schmidt:2012p2518}
H.~J. Schmidt and D.~Singleton, ``Exact radial solution in 2+1 gravity with a
  real scalar field,'' [\eprint{1212.1285}]. 

\bibitem{Kwon:2012p2496}
Y.~Kwon, S.~Nam, and J.-D. Park, ``Extremal Black Holes and Holographic
  C-Theorem,'' [\eprint{1208.4509}].


\bibitem{Radu:2005p2569}
E.~Radu and E.~Winstanley, ``Conformally coupled scalar solitons and black
  holes with negative cosmological constant,'' 
  Phys.Rev. D72 (2005) 024017 [\eprint{gr-qc/0503095}].

\bibitem{Martinez:2002p2565}
C.~Martinez, R.~Troncoso, and J.~Zanelli, ``de Sitter black hole with a
  conformally coupled scalar field in four dimensions,'' 
  Phys.Rev. D67 (2003) 024008 [\eprint{hep-th/0205319}].

\bibitem{Jamil:2012p2563}
M.~Jamil, D.~Momeni, and R.~Myrzakulov, ``Stability of a non-minimally
  conformally coupled scalar field in F(T) cosmology,'' [\eprint{1208.0025}].

\bibitem{Kuriakose:2008p2559}
P.~I. Kuriakose and V.~C. Kuriakose, ``Static Black Hole dressed with a massive
  Scalar field,'' [\eprint{0805.4554}].

\bibitem{Dotti:2007p2555}
G.~Dotti, R.~J. Gleiser, and C.~Martinez, ``Static black hole solutions with a
  self interacting conformally coupled scalar field,'' 
  Phys.Rev.D77:104035,2008 [\eprint{0710.1735}].

\bibitem{Kolyvaris:2011p1518}
T.~Kolyvaris, G.~Koutsoumbas, E.~Papantonopoulos, and G.~Siopsis, ``Einstein
  Hair,'' [\eprint{1111.0263}].

\bibitem{Perez:2013p2534}
A.~Perez, D.~Tempo, and R.~Troncoso, ``Higher spin black hole entropy in three
  dimensions,'' [\eprint{1301.0847}]. 

\bibitem{Anninos:2011p953}
D.~Anninos, T.~Hartman, and A.~Strominger, ``Higher Spin Realization of the
  dS/CFT Correspondence,'' [\eprint{1108.5735}]. 

\bibitem{Ammon:2011p942}
M.~Ammon, M.~Gutperle, P.~Kraus, and E.~Perlmutter, ``Spacetime Geometry in
  Higher Spin Gravity,'' [\eprint{ 1106.4788}]. 
  
\bibitem{Chen:2012p2513}
B.~Chen, J.~Long, and Y.-N. Wang, ``Phase Structure of Higher Spin Black
  Hole,'' [\eprint{1212.6593}]. 
  
\bibitem{David:2012p2251}
J.~R. David, M.~Ferlaino, and S.~P. Kumar, ``Thermodynamics of higher spin
  black holes in 3D,'' [\eprint{1210.0284}].

\bibitem{Chen:2012p2249}
B.~Chen, J.~Long, and Y.-N. Wang, ``Black holes in Truncated Higher Spin
  AdS${}_3$ Gravity,'' [\eprint{1209.6185}]. 
    
\bibitem{Bergshoeff:2009p312}
E.~A. Bergshoeff, O.~Hohm, and P.~K. Townsend, ``Massive Gravity in Three
  Dimensions,'' Phys.Rev.Lett.102:201301,2009 [\eprint{0901.1766}].

\bibitem{Gabadadze:2012p2512}
G.~Gabadadze, G.~Giribet, and A.~Iglesias, ``New Massive Gravity on de Sitter
  Space and Black Holes at the Special Point,'' [\eprint{1212.6279}].

\bibitem{Hohm:2012p2150}
O.~Hohm, A.~Routh, P.~K. Townsend, and B.~Zhang, ``On the Hamiltonian form of
  3D massive gravity,'' [\eprint{1208.0038}]. 

\bibitem{Chen:2012p1989}
B.~Chen, J.~Long, and J.~dong Zhang, ``Classical Aspects of Higher Spin
  Topologically Massive Gravity,'' [\eprint{1204.3282}].

\bibitem{Ohta:2011p1120}
N.~Ohta, ``A Complete Classification of Higher Derivative Gravity in 3D and
  Criticality in 4D,'' [\eprint{1109.4458}].

\bibitem{Bagchi:2011p850}
A.~Bagchi, S.~Lal, A.~Saha, and B.~Sahoo, ``Topologically Massive Higher Spin
  Gravity,'' JHEP 1110 (2011) 150 [\eprint{1107.0915}].

\bibitem{Song:2008p316}
W.~Li, W.~Song and A.~Strominger, ``Chiral Gravity in Three Dimensions,'' 
  JHEP 0804:082,2008 [\eprint{0801.4566}].

\bibitem{Oliva:2012p2290}
J.~Oliva, ``All the solutions of the form M2(warped)xΣ(d-2) for Lovelock
  gravity in vacuum in the Chern-Simons case,'' [\eprint{1210.4123}]. 

\bibitem{Brihaye:2010p17}
Y.~Brihaye, E.~Radu, and D.~H. Tchrakian, ``Asymptotically flat, stable black
  hole solutions in Einstein-Yang-Mills-Chern-Simons theory,'' 
  Phys.Rev.Lett.106:071101,2011 [\eprint{1011.1624}].

\bibitem{Moore:1989yh} 
  G.~W.~Moore and N.~Seiberg,
  ``Taming the Conformal Zoo,''  Phys.\ Lett.\ B {\bf 220}, 422 (1989). 

\bibitem{Elitzur:1989nr} 
  S.~Elitzur, G.~W.~Moore, A.~Schwimmer and N.~Seiberg,
  ``Remarks on the Canonical Quantization of the Chern-Simons-Witten Theory,'' 
   Nucl.\ Phys.\ B {\bf 326}, 108 (1989). 

\bibitem{Brihaye:2011p2541}
Y.~Brihaye and B.~Hartmann, ``A scalar field instability of rotating and
  charged black holes in (4+1)-dimensional Anti-de Sitter 
  space-time,'' [\eprint{1112.6315}].

\bibitem{Anabalon:2009p2560}
A.~Anabalon and H.~Maeda, ``New Charged Black Holes with Conformal Scalar
  Hair,'' Phys.Rev. D81 (2010) 041501 [\eprint{0907.0219}].

\bibitem{Anabalon:2012p2557}
A.~Anabalon, ``Exact Hairy Black Holes,'' [\eprint{1211.2765}].

\bibitem{Anabalon:2012p2562}
A.~Anabalon and A.~Cisterna, ``Asymptotically (anti) de Sitter Black Holes 
 and Wormholes with a Self Interacting Scalar Field in Four  
 Dimensions,'' [\eprint{1201.2008}].

\bibitem{Charmousis:2009p2558}
C.~Charmousis, T.~Kolyvaris, and E.~Papantonopoulos, ``Charged C-metric with
  conformally coupled scalar field,'' 
  Class.Quant.Grav.26:175012,2009 [\eprint{0906.5568}].

\bibitem{Martinez:2005p2568}
C.~Martinez, J.~P. Staforelli, and R.~Troncoso, ``Topological black holes
  dressed with a conformally coupled scalar field and electric charge,'' 
  Phys.Rev.D74:044028,2006 [\eprint{hep-th/0512022}].

\bibitem{Nucamendi:2003p2566}
U.~Nucamendi and M.~Salgado, ``Scalar hairy black holes and solitons in
  asymptotically flat spacetimes,'' Phys.Rev. D68 (2003) 044026 
  [\eprint{gr-qc/0301062}].

\bibitem{Ida}
  D.~Ida,
  ``No black hole theorem in three-dimensional gravity,''
  Phys. Rev. Lett. 85 (2000) 3758 [\eprint{gr-qc/0005129}].


\bibitem{Garcia:2003p2522}
A.~Garcia, F.~W. Hehl, C.~Heinicke, and A.~Macias, ``The Cotton tensor in
  Riemannian spacetimes,'' Class.Quant.Grav. 21 (2004) 1099-1118 
  [\eprint{gr-qc/0309008}].




\end{thebibliography}

\providecommand{\href}[2]{#2}\begingroup
\footnotesize\itemsep=0pt
\providecommand{\eprint}[2][]{\href{http://arxiv.org/abs/#2}{arXiv:#2}}

\endgroup

\end{document}